# Spontaneous Economic Order[†]

Yong Tao[‡]

School of Economics and Business Administration,
Chongqing University,
Chongqing 400044, China

**Abstract**

This paper provides an attempt to formalize Hayek's notion of spontaneous order within the framework of the Arrow-Debreu economy. Our study shows that if a competitive economy is enough fair and free, then a spontaneous economic order shall emerge in long-run competitive equilibria so that social members together occupy an optimal distribution of income. Despite this, the spontaneous order might degenerate in the form of economic crises whenever an equilibrium economy approaches the extreme competition. Remarkably, such a theoretical framework of spontaneous order provides a bridge linking Austrian economics and Neoclassical economics, where we shall comprehend a truth: "Freedom promotes technological progress".

[†] Project supported by the Scholarship Award for Excellent Doctoral Student Granted by Ministry of Education of China (2012), (Grant No. 0903005109081-019).
[‡] Corresponding author.
   E-mail address: taoyingyong@yahoo.com

**Dedicated to My Mother**

# 1. Introduction

*Spontaneous order* in economic interactions presented by Hayek (1948) is an important notion for economics. It, originates from the interactions of members of society, is something to which everyone contributes, from which everyone benefits, which everyone normally takes for granted, but which individuals rarely understand (Witt, 1997). Hayek believed that if the degree of freedom of a society, under the constraint of limited resources, reaches maximum, then a spontaneous order would emerge so that social members together occupy an optimal allocation of resources. Nevertheless, nobody makes clear what is the spontaneous order, for example, is it a natural law? On the other hand, we have known that in his early years Hayek focused his research on the theory of business cycle and later turned to the theory of a spontaneous economic order (Witt, 1997). Hence we want to ask a question: "Are there some relationships between economic crisis and spontaneous order?" Unfortunately, Hayek never reconsidered business cycle theory in the light of his later thought.

We are all now witnesses of the present huge financial crisis started in 2008. As for this crisis, many people attribute the origin of it to the laissez faire policies supporting free markets (Bouchaud, 2008). Therefore, a natural question should be answered: "Is the spontaneous order really valid for free economies?" Regrettably, we can not answer this question because so far there is no a general equilibrium model which is really on the basis of the principle of spontaneous order[1]. For example, the standard model of modern economics is called *Dynamic Stochastic General Equilibrium* (DSGE) in which there is no an interesting variable corresponding to spontaneous order or degree of freedom of economic systems. Often, we do not find a strict solution to DSGE, but we can prove the existence of the equilibrium solution which has some fine properties such that economic crises are ruled out. This means, we can not make clear the origin of economic crises only via the general equilibrium theory. As a result, some scholars appealed to abandon the DSGE and *Neoclassical economics* so as to develop some new economic theories of early warning of economic crises, see Hodgson (2009), Leijonhufvud (2009) and Farmer et al (2009). However, it is illogical to abandon Neoclassical economics, since it had obtained great success in the past. Later, we shall see that the spontaneous order is an inevitable outcome when *multiple* competitive equilibria arise, and that the economic crisis is an *unstable* state of spontaneous order.

---

[1] It is worth mentioning that there have been many literatures in which some authors try to connect the principle of spontaneous order and the method of evolutionary game, e.g., see Schotter (1981), Sugden (1989) and Young (1993) (1996). Nonetheless, these excellent attempts pay more attentions to the order of social rules (e.g., conventions or institutions) rather than the order of economic rules (e.g., distribution of wealth or income). Obviously, the imbalance of the latter is more likely associated with economic crises. And the latter which we are chiefly concerned with is easier to be tested empirically.

As is well known, to guarantee that a competitive economy has a unique equilibrium outcome, we must assume that each consumer's preferences are strictly convex (Jehle and Reny, 2001; Page 188) and meanwhile that each firm's production possibility sets are strongly convex (Jehle and Reny, 2001; Page 206). Indeed, strict convexity of preferences is necessary (for Neoclassical economics), since it is exhibiting the principle of diminishing marginal rate of substitution in consumption (Jehle and Reny, 2001; Page 12). However, strong convexity will rule out constant returns to scale in production. And the latter is considerably important in the Neoclassical production theory. If, instead, merely convexity of production possibility sets is assumed, the existence of equilibrium outcome can still be proved (Debreu, 1971; Page 84). It is crucial that the convexity (rather than strong convexity) of production sets allows the possibility of constant returns to scale for firms (Jehle and Reny, 2001; Page 216). So, if one notes that the constant returns technology is a (only sensible) long-run production technology (Varian, 1992; Page 356), the convexity of production possibility sets actually ensures the existence of long-run equilibrium outcome. Furthermore, if one notices that the long-run level of profits for a competitive firm that has constant returns to scale is a zero level of profits (Varian, 2003; Page 340), then one shall have to face an interesting (long-run) equilibrium state: each firm always gains zero economic profit no matter how it behaves. This strongly implies that the long-run competitive economy may have multiple or indeterminate equilibrium outcomes.

To strictly make clear whether or not a long-run competitive economy produces multiple equilibria, we need to introduce an exact definition for such an economy. In this paper, a long-run competitive economy would be specified by an Arrow-Debreu economy with additivity and publicly available technology. Traditional literature (Mas-Collel et al, 1995; Page 334) demonstrated that the long-run competitive economy is a situation of competitive economies when free entry is permitted. Roughly speaking, additivity means that there is free entry for firms into a possible industry (Debreu, 1971; Page 41). Hence, additivity and publicly available technology would together guarantee that there is free entry for firms into any (technology) industry. Moreover, publicly available technology (Mas-Collel et al, 1995; Page 653) implies that firms produce their products with a similar or an identical technology; therefore, monopolistic competition and perfect competition are allowed for as well. Because of these above, we believe that an Arrow-Debreu economy with additivity and publicly available technology exactly describes the long-run competitive economy. Later, we shall prove that such an economy indeed has many (even infinitely many) equilibrium outcomes. Not only that, according to the first fundamental theorem of Welfare economics, these equilibrium outcomes should be all Pareto optimal. Because each equilibrium outcome is associated with a different social state, multiple equilibrium outcomes actually imply an uncertain economic world. To eliminate the uncertainty, welfare economists attempt to search the best outcome through an imaginary social welfare function. However, *Arrow's Impossibility Theorem* had refuted this attempt (Jehle and Reny, 2001; Page 243). This means, within the framework of Neoclassical and Welfare economics, a long-run competitive economy does produce an uncertain economic world.

In this paper, we propose a new scheme for "eliminating" the uncertainty. Our plan is to extend the theoretical framework of Neoclassical economics so as to exhibit the principle of spontaneous order, whence we shall find the economic rule hidden behind multiple equilibrium outcomes. The intuition behind our approach is as follows. Now that a long-run competitive economy produces multiple equilibrium outcomes, each of which is Pareto optimal. We can assume that all these outcomes (or corresponding social states) are equally likely to occur (or equivalently, to be selected with equal opportunities as collective decisions). Equal opportunities among equilibrium outcomes essentially imply an absolutely fair world in which there is no any difference between all the outcomes. Then, if we can find an economic order (or a convention) that contains the most equilibrium outcomes, it does occur with the highest probability (compared to other economic orders). We shall define such an economic order with the highest probability as the spontaneous economic order. Undoubtedly, such a definition exhibits Hayek's core idea (Sugden, 1989): Spontaneous order is a convention which is *most likely* to evolve and *survive*.

To formalize the intuition above, we must first prove that a long-run competitive economy does produce a set of equilibrium outcomes, which is denoted, for example, by $B = \{B_1, B_2, B_3, B_4\}$. Meanwhile we must show that every element in the set $B$ is Pareto optimal. In this case, if the competitive economy is absolutely fair[2], we can think of it as a fair procedure which would translate its fairness to the outcomes (Rawls, 1999; Page 75) so that all the social members would be indifferent between these elements; that is,

$$B_1 \sim B_2 \sim B_3 \sim B_4 \qquad (1.1)$$

Because all of these equilibrium outcomes are equally fair, every social member will have no desire of opposing or preferring a certain outcome. Then every equilibrium outcome should occur with an equal probability (or equivalently, every equilibrium outcome should be selected with an equal opportunity as collective decisions). For example, (1.1) implies that every equilibrium outcome $B_i$ should occur with the probability $\frac{1}{4}$, where, $i = 1,2,3,4$. Assume now that these four equilibrium outcomes could be divided into the following three economic orders[3] (or three conventions): $a_1 = \{B_1\}$, $a_2 = \{B_2, B_3\}$ and $a_3 = \{B_4\}$, then we have to conclude that $a_1$ occurs with the probability $\frac{1}{4}$, and that $a_2$ occurs with the probability $\frac{1}{2}$, and that $a_3$ occurs with the probability $\frac{1}{4}$. Undoubtedly, $a_2$ will occur with the highest probability, this is because $a_2$ contains the most equilibrium outcomes. Normally, more equilibrium outcomes imply more choice opportunities or greater opportunity-freedom, see Sen (1993). From this meaning, $a_2$ is an economic order not only with fairness but also with the greatest opportunity-freedom. And $a_2$ is of course a spontaneous economic order according to previous definition.

---

[2] It is worth mentioning that there might be difficulty concerning the possibility of satisfying fairness and Pareto optimality objectives simultaneously when interpersonal comparisons of utility are allowed (Pazner and Schmeidler, 1974). However, one can rule out this difficulty by insisting on the standpoint of ordinal utility (Pazner and Schmeidler, 1978).

[3] $a_1 = \{B_1\}$ represents an economic order or a convention that allows equilibrium outcome $B_1$ to occur. Likewise, $a_2 = \{B_2, B_3\}$ allows $B_2$ and $B_3$; $a_3 = \{B_4\}$ allows $B_4$.

Summarizing the analyses above, we are able to develop three steps for seeking the spontaneous economic order. First, try to find all possible equilibrium outcomes of a competitive economy. Second, divide all these equilibrium outcomes into some economic orders. Finally, find the economic order with the highest probability through the normative criteria about fairness and freedom.

The main purpose of this paper is to seek the spontaneous order of a *long-run* competitive economy using the three steps above. To arrive at this purpose, with each economic order we shall associate a possible individuals' revenue distribution. With this setting, we later show that the spontaneous order of a monopolistic-competitive economy will obey a stable rule: Boltzmann distribution; and that the spontaneous order of a perfectly competitive economy will obey an unstable rule: Bose-Einstein distribution. And the instability of the latter might cause economic crises (Tao, 2010). Interestingly, some recent empirical investigations have confirmed that the individuals' revenue distribution of free economies (e.g. USA) during the period of stable economy obeys Boltzmann distribution, see Yakovenko & Rosser (2009), Clementi, et al (2012); and obeys Bose-Einstein distribution in the run-up to an economic crisis, see Kürten & Kusmartsev (2011), Kusmartsev (2011).

Since our finding might appear somewhat surprising, we try to convey an intuition for the result. As is well known, in microeconomics, there are only four types of markets: perfectly competitive market, monopolistic-competitive market, oligopoly market and perfectly monopoly market. Also, for these four types of markets, the perfectly competitive market is of course most efficient. It is worth mentioning that, before every extremely serious economic crisis occurred, there always had, without exception, appeared the extremely prosperous economies, especially in financial market[4]. Therefore, a natural question arises: which type of the above four markets shall, most likely, cause the extremely prosperous economy?

---

[4] Interestingly, compared to all the other real markets, the financial market is closest to a perfectly competitive market. This is the reason why the Black-Scholes equation of option pricing can be well applied in a financial market. The starting point of the Black-Scholes equation of option pricing is that the change in the price of stock obeys the law of Brownian movement. Only the perfectly competitive market, which is free of monopolization, is closest to such an ideal state. Minsky (1986) ever claimed that the finance was the cause of the instability of capitalism. Now, according to our theory, that is because the financial market is closest to perfect competition.

Logically, the answer should be the most efficient market-namely, perfectly competitive market.

Let us recall that there was a common feature emerging in the past three serious economic crises[5]; that is, before they occurred, there, without exception, had appeared the extremely prosperous economies. Of course, aimed at every past economic crisis, someone might always find an explanation which seems right for origin of this crisis, e.g. asymmetric information, currency mismatch between assets and liabilities of firms (Deesomsak et al, 2009), even greedy (selfish) for explaining the origin of economic crisis in 2008. However, we need to remind that the selfish is just one of several axioms of economics. We have known that perfect competition is the extreme case of competitive economies. Logically, as the competition in a free economy increases (for example, long-term policy of low rates of interest started in 2001 stimulated the competition in USA economy), the economy shall naturally evolve toward extreme competition (i.e., perfect competition) from monopolistic competition; however, according to our theory, perfect competition is not stable, and may cause economic crisis. That is to say, a thing turns into its opposite if pushed too far.

The organization of our paper is as follows. Section 2 introduces the definition of long-run competitive equilibrium within the framework of Arrow-Debreu economy. Subsections 3.1 and 3.2 prove that a long-run competitive economy has at least an equilibrium outcome. Subsection 3.3 and section 4 show that a long-run equilibrium outcome might generate many (even infinite many) long-run equilibrium outcomes. Section 5 introduces the concepts of economic order and spontaneous economic order, and meanwhile shows that all the long-run equilibrium outcomes can be appropriately (non-repeated) assigned into some economic orders. Section 6 shows that one can seek the spontaneous economic order through some normative criteria about fairness and freedom. Section 7 investigates the possible link between spontaneous economic order and Neoclassical macroeconomics, and meanwhile introduces some empirical evidences supporting our results. Section 8 explores the relationship between technological progress and social freedom. In section 9, our conclusion follows.

## 2. Preliminaries

---

[5] These three economic crises are respectively: Great Depression in 1929, Asian financial crises in 1997, and American subprime crisis in 2008.

We begin by describing a competitive economy which is composed of a vast number of agents (consumers and firms) and diverse industries. Following the standard framework of the Neoclassical economics (Mas-Collel et al, 1995; Page 579), we assume that there are $M$ consumers, $N$ firms and $L$ commodities. Every consumer $i=1,...,M$ is specified by a consumption set $X_i \subset R^L$, a preference relation $\succsim_i$ on $X_i$, an initial endowment vector $\omega_i \in R^L$, and an ownership share $\theta_{ij} \geq 0$ of each firm $j=1,...,N$ (where $\sum_{i=1}^{M} \theta_{ij} = 1$). Each firm $j$ is characterized by a production set $Y_j \subset R^L$. All allocations for such an economy is a collection of consumption and production vectors:

$$((x),(y)) = (x_1,..., x_M, y_1,..., y_N) \in X_1 \times ... \times X_M \times Y_1 \times ... \times Y_N,$$

where $x_i = (x_{1i},..., x_{Li})$ and $y_j = (y_{1j},..., y_{Lj})$.

**2.1 Arrow-Debreu economy and competitive equilibrium**

A well-known definition for competitive equilibrium is introduced as follow.

***Definition 2.1*** (Mas-Collel et al, 1995; Page 579)**:** An allocation $((x^c),(y^c))$ and a price vector $p = (p_1,..., p_L)$ constitute a competitive (or Walrasian) equilibrium if the following three conditions are satisfied:

(1). Profit maximization: For every firm $j$, $y_j^c \in Y_j$ maximizes profits in $Y_j$; that is,

$$p \cdot y_j \leq p \cdot y_j^c \text{ for all } y_j \in Y_j$$

(2). Utility maximization: For every consumer $i$, $x_i^c \in X_i$ is maximal for $\succsim_i$ in the budget set:

$$\left\{ x_i \in X_i : p \cdot x_i \leq p \cdot \omega_i + \sum_{j=1}^{N} \theta_{ij} p \cdot y_j^c \right\}.$$

(3). Market clearing: $\sum_{i=1}^{M} x_i^c = \sum_{i=1}^{M} \omega_i + \sum_{j=1}^{N} y_j^c$.

One can verify that the equilibrium allocation $((x^c),(y^c))$ does exist if the following nine conditions are satisfied (Debreu, 1971; page 84).

For every consumer $i$:

(a). Each consumer's consumption set $X_i$ is closed, convex, and bounded below;

(b). There is no satiation consumption bundle for any consumer;

(c). For each consumer $i = 1,...,M$, the sets $\left\{ x_i \in X_i \middle| x_i \succsim_i x_i' \right\}$ and $\left\{ x_i \in X_i \middle| x_i' \succsim_i x_i \right\}$ are closed;

(d). If $x_i^1$ and $x_i^2$ are two points of $X_i$ and if $t$ is a real number in $(0,1)$, then $x_i^2 \succ_i x_i^1$ implies $tx_i^2 + (1-t)x_i^1 \succ_i x_i^1$;

(e). There is $x_i^0$ in $X_i$, such that $x_i^0 \ll \omega_i$;

For every firm $j$:

(f). $0 \in Y_j$;

(g). $Y = \sum_{j=1}^{N} Y_j$ is closed and convex;

(h). $-Y \cap Y = \{0\}$;

(i).  $-R_+^L \subset Y$.

If a competitive economy satisfies (a)-(i), it is called the Arrow-Debreu economy (Arrow and Debreu, 1954). In particular, (d) will guarantee that each consumer's preferences are *strictly convex*. It should be noted that Debreu did not emphasize the strict convexity of preferences. Instead, merely convexity of preferences was assumed in his famous book (Debreu, 1971; Page 84). However, strict convexity of preferences is necessary for Neoclassical economics, since it is exhibiting the principle of diminishing marginal rate of substitution in consumption (Jehle and Reny, 2001; Page 12). Technically, strict convexity of preferences will guarantee that $(x_1^c,...,x_M^c)$ is a unique equilibrium consumption allocation.

Here, we do not assume the strong convexity of production possibility sets, since it will rule out constant returns to scale in production (Jehle and Reny, 2001; Page 206). Later, we shall see that constant returns technology is necessary when long-run competition is taken into account, and that (f)-(i) allow the possibility of constant returns to scale for firms.

## 2.2 Long-run competitive equilibrium

Following Mas-Collel et al (1995; Page 334), we consider the case in which "an infinite number of firms can potentially be formed"; that is, $N \to \infty$. Moreover, each firm has access to the publicly available technology so that it might enter and exit an industry in response to profit opportunities. "This scenario, known as a situation of free entry, is often a reasonable approximation when we think of long-run outcomes" in an industry (or a market). Under such a scenario, Mas-Collel et al deduced (1995; Page 335): "A firm will enter the market if it can earn positive profits at the going market price and will exit if it can make only negative profits at any positive production level given this price. If all firms, active and potential, take prices as unaffected by their own actions, this implies that active firms must earn exactly *zero profits* in any long-run competitive equilibrium; otherwise, we would have either no firms willing to be active in the market (if profits were negative) or an infinite number of firms entering the market (if profits were positive)". From this reasoning, if all the industries (or markets) stay at long-run competitive equilibria, then we do have:

$p \cdot y_j^c = 0$,                     (2.1)

$j = 1,..N$,

Combining the Definition 2.1 and (2.1), we can present a natural definition for long-run competitive equilibrium as follow.

***Definition 2.2:*** An allocation $((x^*),(y^*))$ and a price vector $p = (p_1,..., p_L)$ constitute a long-run competitive equilibrium if the following three conditions are satisfied:

(1). For every firm $j$, there exists $y_j^* \in Y_j$ such that $p \cdot y_j \leq p \cdot y_j^* = 0$ for all $y_j \in Y_j$.

(2). For every consumer $i$, $x_i^* \in X_i$ is maximal for $\succsim_i$ in the budget set: $\{x_i \in X_i : p \cdot x_i \leq p \cdot \omega_i\}$.

(3). $\sum_{i=1}^{M} x_i^* = \sum_{i=1}^{M} \omega_i + \sum_{j=1}^{N} y_j^*$.

Obviously, in contrast to the Definition 2.1, the Definition 2.2 has a stronger constraint; that is, the maximum profit of every firm $j$, $p \cdot y_j^*$, is restricted to be null. Because of this, we can not guarantee that $((x^*),(y^*))$ does exist even if (a)-(i) are satisfied. In the next section, our attention will be focused on the existence of a long-run equilibrium allocation $((x^*),(y^*))$. To avoid confusion, when we mention an equilibrium allocation in the subsequent sections, we always mean that it denotes a long-run equilibrium allocation.

## 3. Long-run competitive economy

### 3.1 Assumptions

Now we explore what conditions will restrict the maximum profit of every firm $j$ to be null within the framework of Arrow-Debreu economy.

***Assumption 3.1*** (additivity)**:** $Y_j + Y_j \subset Y_j$ for every $j$.

If the production set of the $j$ th firm, $Y_j$, can be interpreted as an industry, the additivity Assumption 3.1 means that there is free entry for firms into that industry (Debreu, 1971; Page 41). More importantly, one has the result as below:

***Theorem 3.1*:** If Assumption 3.1 and (f) are satisfied, then the maximum profit of every firm $j$ is zero; that is, $p \cdot y_j^* = 0$ for $j = 1,...,N$.

***Proof***. See page 45 in Debreu (1971). □

The Theorem 3.1 demonstrates that the Arrow-Debreu economy under Assumption 3.1 will restrict the maximum profit of every firm to be null (if the maximum profit exists). Despite this, the additivity Assumption 3.1 does not really imply free entry, since $Y_j$ represents a private production set (of the $j$ th firm) rather than a public industry. However, if the following assumption is satisfied, then the additivity Assumption 3.1 will imply free entry.

***Assumption 3.2*** (publicly available technology[6])**:** $Y_1 = Y_2 = ... = Y_N$.

The Assumption 3.2 implies that every firm has free access to one another's technology. Then every $Y_j$ represents a public production set, and thereby can be interpreted as a public (open) industry. Because of this, Assumptions 3.1 and 3.2 together imply free entry. More generally, we have:

***Proposition 3.1*:** Assumptions 3.1 and 3.2 together guarantee that $Y_j = Y$ for $j = 1,2,...,N$ and thereby that $Y + Y \subset Y$, where $Y = \sum_{j=1}^{N} Y_j$.

---

[6] Publicly available technology coincides with Rawls' principle of fair equality of opportunity (Rawls, 1999; Page 63)

***Proof***. To verify this proposition, we only need to prove that for any $y \in Y$ there has to be $y \in Y_j$. Because $y \in Y$, we have $y = \sum_{k=1}^{N} y_k$, where $y_k \in Y_k$. Then, by Assumption 3.2, we immediately arrive at $y_k \in Y_j$ for $k = 1,2,...N$, where $j = 1,2,...,N$. Finally, by Assumption 3.1, we have $y = \sum_{k=1}^{N} y_k \in Y_j$. □

### 3.2 Existence of long-run competitive equilibrium

Because the Arrow-Debreu economy under Assumption 3.1 will restrict the maximum profit of every firm to be null (if the maximum profit exists), and also because Assumptions 3.1 and 3.2 together imply free entry, whence we note that if the Arrow-Debreu economy under Assumptions 3.1 and 3.2 has an equilibrium allocation $((x'),(y'))$, then $((x'),(y'))$ will satisfy the Definition 2.2 automatically. Therefore, we can introduce an exact definition for long-run competitive economy as follow.

***Definition 3.1***: A competitive economy is a long-run competitive economy (LRCE) if and only if:
(1). (a)-(i) are satisfied;
(2). Assumptions 3.1 and 3.2 hold.

To verify that a LRCE has at least an equilibrium allocation, we only need to prove that (a)-(i) are compatible with Assumptions 3.1 and 3.2. This is because (a)-(i) themselves would ensure the existence of an equilibrium allocation (Debreu,1971; page 84). Before proceeding to prove this, let us introduce two lemmas.

***Lemma 3.1***: If (f)-(g) are satisfied and if Assumptions 3.1 and 3.2 hold, then $Y$ is a cone with vertex 0, i.e., $y \in Y$ implies $ty \in Y$ for any scalar $t \geq 0$.

***Proof***. First, by (f) one has $0 \in Y$, and by (g) $Y$ satisfies convexity; therefore, for any $y \in Y$ and any $c \in [0,1]$, one has $cy = cy + (1-c) \cdot 0 \in Y$. Second, by proposition 3.1 (because Assumptions 3.1 and 3.2 hold) $Y$ satisfies additivity; that is, for any non-negative integer $k$, one has $ky \in Y$. Let $t$ be any non-negative number satisfying $t \leq k$, then the two preparations above imply $ty = \frac{t}{k} \cdot ky \in Y$. □

Recall that a cone with vertex 0 implies constant returns to scale (Debreu,1971; page 46), we immediately have two corollaries:

***Corollary 3.1***: The production set $Y$ exhibits constant returns to scale.

***Corollary 3.2***: The production set of each firm, $Y_j$, exhibits constant returns to scale.

***Lemma 3.2***: If $Y$ is a cone with vertex 0 and meanwhile if $Y$ is closed and convex, then $Y$ must be a closed, convex cone with vertex 0.

***Proof***. See page 42 in Debreu (1971) □

Combining the Lemmas 3.1 and 3.2 we have:

***Theorem 3.2***: A LRCE has at least an equilibrium allocation $((x^*),(y^*))$.

***Proof***. We now proceed to prove that (a)-(i) are compatible with Assumptions 3.1 and 3.2. By Lemma 3.1, (f)-(g) together with Assumptions 3.1 and 3.2 guarantee that $Y$ is a cone with vertex 0. Then by (g) and Lemma 3.2, $Y$ is further a closed, convex cone with vertex 0. Undoubtedly, such a result does not contradict (f)-(g). This means that (a)-(i) might be still satisfied even if Assumptions 3.1 and 3.2 hold. □

Now that the LRCE satisfies (a)-(i) (see the proof above), we immediately arrive at two corollaries as follows:

***Corollary 3.3***: The LRCE is an Arrow-Debreu economy.

***Corollary 3.4***: Any long-run equilibrium allocation $((x^*),(y^*))$ is a competitive (or Walrasian) equilibrium.

### 3.3 Multiplicity of long-run competitive equilibria

In subsection 3.2, we have proved that a LRCE has at least an equilibrium allocation $((x^*),(y^*)) = (x_1^*,..., x_M^*, y_1^*,..., y_N^*)$. Next, we show that $((x^*),(y^*))$ might generate many (even infinitely many) equilibrium allocations.

Let $z(p) = \sum_{j=1}^{N} y_j^*$ denote the aggregate production vector, then we have:

***Proposition 3.2***: $z(p) \in Y_j$ for $j = 1,..., N$.

***Proof***. By Proposition 3.1 one has $z(p) \in Y = Y_j$ for $j = 1,..., N$. □

***Lemma 3.3***: $tz(p) \in Y_j$ for $j = 1,..., N$, where $t \geq 0$.

***Proof***. By Lemmas 3.1 and Proposition 3.2 one concludes $tz(p) \in Y_j$. □

***Lemma 3.4***: $p \cdot z(p) = 0$.

***Proof***. By Definition 2.2 one has $p \cdot y_j^* = 0$ for $j = 1,..., N$. □

Let us consider a sequence of numbers, $\{t_j\}_{j=1}^{N}$, satisfying:

$$\begin{cases} t_j \geq 0 \quad for\ j = 1, 2, ...N, \\ \sum_{j=1}^{N} t_j = 1 \end{cases} \quad (3.1)$$

Then, by Lemmas 3.3 and 3.4 we can prove the following proposition.

***Proposition 3.3***: Let

$$y_j^{'}(t_j) = t_j z(p) \quad (3.2)$$

for $j = 1, 2, ..., N$, then $(x_1^*, ..., x_M^*, y_1^{'}(t_1), ..., y_N^{'}(t_N))$ constitutes a long-run equilibrium allocation.

***Proof***. Because $\sum_{j=1}^{N} y_j^{'}(t_j) = z(p) = \sum_{j=1}^{n} y_j^*$, we only need to verify that each $y_j^{'}(t_j)$ satisfies the condition (1) in Definition 2.2. To this end, by Lemma 3.3 $y_j^{'}(t_j) \in Y_j$ and by Lemma 3.4 $p \cdot y_j^{'}(t_j) = 0$, where $j = 1, ..., N$. □

The proof above implies two corollaries as below:

***Corollary 3.5***: $\sum_{j=1}^{N} y_j^{'}(t_j) = z(p)$.

***Corollary 3.6***: $p \cdot y_j^{'}(t_j) = 0$ for $j = 1, ..., N$.

The Proposition 3.3 indicates that each sequence $\{t_j\}_{j=1}^{N}$ satisfying (3.1) will produce a different long-run equilibrium allocation correspondingly. Undoubtedly, there might be infinitely many possible sequences $\{t_j\}_{j=1}^{N}$ satisfying (3.1), so will be long-run equilibrium allocations.

***Lemma 3.5*** (First fundamental theorem of Welfare economics)**:** Any Walrasian equilibrium allocation is Pareto optimal.

***Proof***. See page 549 in Mas-Collel et al (1995). □

Using Corollary 3.4 and Lemma 3.5 we can prove an important proposition.

***Proposition 3.4***: Any equilibrium allocation $\left(x_1^*,..., x_M^*, y_1^{'}(t_1),..., y_N^{'}(t_N)\right)$ obeying (3.1) is Pareto optimal.

*Proof*. By Proposition 3.3 and Corollary 3.4 one concludes that any $\left(x_1^*,..., x_M^*, y_1^{'}(t_1),..., y_N^{'}(t_N)\right)$ obeying (3.1) is a Walrasian equilibrium. Then by Lemma 3.5 we complete this proof. □

## 4. Uncertainty of social choice

The Propositions 3.3 and 3.4 together demonstrate that an equilibrium allocation $\left(x_1^*,...,x_M^*, y_1^*,..., y_N^*\right)$ might generate uncertain equilibrium outcomes $\left(x_1^*,..., x_M^*, y_1^{'}(t_1),..., y_N^{'}(t_N)\right)$, each of which is Pareto optimal. And the following proposition will further reveal that the uncertainty of equilibrium outcomes is due to production side rather than consumption side.

***Proposition 4.1***: $\left(x_1^*, x_2^*,...,x_M^*\right)$ is a unique equilibrium consumption allocation.

*Proof.* If there were another equilibrium consumption allocation $\left(x_1^{'}, x_2^{'},...,x_M^{'}\right)$ satisfying $x_i^{'} \succeq_{\sim i} x_i^*$ for $i=1,...,M$, then by the strict convexity condition (d) we do have $x_i^{''} = tx_i^{'} + (1-t)x_i^* \in X_i$ so that $x_i^{''} \succ_i x_i^*$ contradicting the condition (2) of Definition 2.2, where $0<t<1$. □

Obviously, using the Proposition 4.1 we immediately arrive at two corollaries:

***Corollary 4.1***: $\left(x_1^*,..., x_M^*, y_1^{'}(t_1),..., y_N^{'}(t_N)\right)$ can be reduced to $\left(y_1^{'}(t_1),..., y_N^{'}(t_N)\right)$.

***Corollary 4.2***: $z(p)$ is a fixed vector.

Proposition 4.1 and Corollary 4.1 imply that all equilibria involve the same consumption vector, hence all consumers are indifferent between all equilibria and the multiplicity arises only from the distribution of production. Then a doubt might occur: Since the publicly available technology is assumed (see Assumption 3.2), the multiplicity is perhaps a meaningless (or spurious) multiplicity. However, the multiplicity of equilibria must be admitted, since the *economic crises* are just hidden in such a multiplicity. To see this, we consider a possible *long-run* equilibrium outcome $\left(x_1^*,...,x_M^*,z(p),0,...,0\right)$, where $y_1^{'}(t_1)=z(p)$ and $y_j^{'}(t_j)=0$ for $j=2,...,N$. This equilibrium itself strongly indicates an economic crisis: Only one firm survives, and others all go bankrupt since they stop production for a long time. Later we shall see that such an equilibrium involving economic crisis will not occur in a monopolistic-competitive economy[7]. Nevertheless, we can not rule out it in a perfectly competitive economy[8].

Moreover, the Corollary 4.1 reminds us: To explore the uncertainty of equilibrium outcomes, we only need to analyze the equilibrium production allocation $\left(y_1^{'}(t_1),...,y_N^{'}(t_N)\right)$. For convenience, we might as well denote by $\left(y_1^{'}(t_1),...,y_N^{'}(t_N)\right)$ the long-run equilibrium allocation.

**4.1 Equilibrium revenue allocation**

Without loss of generality, we assume that $z(p)$ has at most one positive component (namely, a single output[9]).

---

[7] (7.9) shows that firms' revenue (or equivalently "output value") distribution in a monopolistic-competitive economy obeys the exponential law. Then there is no possibility that one firm's output value is positive, and others' all are null.

[8] (7.9) shows that firms' revenue (or equivalently "output value") distribution in a perfectly competitive economy is unstable, since the denominator of (7.9) corresponding to $I=1$ may equal zero. Then there is indeed a possibility that one firm's output value is positive, and others' all are null. More details see Tao (2010).

[9] Indeed, the assumption about single output seems very restrictive; however, it made only to keep our writing to follow succinct. And this assumption will not affect the universality of our results.

***Assumption 4.1*:**

$$\begin{cases} z(p) = (z_1(p), \ldots z_m(p), \ldots z_L(p)) \\ \quad\quad z_m(p) \geq 0 \\ z_l(p) \leq 0, \quad l = 1, \ldots m-1, m+1, \ldots L \end{cases}, \quad (4.1)$$

where, $z_m(p)$ stands for outputs' amount and $z_l(p)$ stands for inputs' amount.

Substituting (4.1) into (3.2) we conclude that the equilibrium outputs' amount of the $j$ th firm is specified by $t_j z_m(p)$. Because the equilibrium price of the $m$ th commodity (i.e., output) is denoted by $p_m$, whence the $j$ th firm will obtain $t_j p_m z_m(p)$ units of revenue.

If we refer to $\varepsilon_j(t_j)$ as the *equilibrium revenue* of the $j$ th firm, and refer to $\Pi$ as the *equilibrium total revenue* of an economy, then we have:

$$\varepsilon_j(t_j) = t_j p_m z_m(p), \quad (4.2)$$

$$\sum_{j=1}^{N} \varepsilon_j(t_j) = \Pi. \quad (4.3)$$

Substituting (4.2) into (4.3) we arrive at:

$$\Pi = p_m z_m(p). \quad (4.4)$$

Using (4.4), (4.2) can be rewritten as:

$$\varepsilon_j(t_j) = t_j \Pi. \quad (4.5)$$

***Definition 4.1*:** An equilibrium revenue allocation is a collection of firms' revenue scalars:

$$(\varepsilon_1(t_1), \ldots \varepsilon_N(t_N)). \quad (4.6)$$

Obviously, an equilibrium revenue allocation $(\varepsilon_1(t_1), \ldots, \varepsilon_N(t_N))$ shows revenue allocations among $N$ firms when an economy arrives in long-run competitive equilibria, so there is no essential distinction[10] in denoting either by $(\varepsilon_1(t_1),...,\varepsilon_N(t_N))$ or by $(y_1'(t_1),...,y_N'(t_N))$ the long-run equilibrium allocation. In reality, however, compared to a production allocation $(y_1'(t_1),...,y_N'(t_N))$, a revenue allocation $(\varepsilon_1(t_1),...,\varepsilon_N(t_N))$ is often easier to be tested empirically. Because of this, we are more interested in exploring the revenue allocation of an economy. In the remainder of this paper, we always denote by $(\varepsilon_1(t_1),...,\varepsilon_N(t_N))$ the long-run equilibrium allocation (or equilibrium outcome).

### 4.2 Uncertain equilibrium outcomes

Combining (3.1) and (4.5) one easily notes that any revenue allocation $(\varepsilon_1(t_1),...,\varepsilon_N(t_N))$ satisfying the following requirements:

$$\begin{cases} \varepsilon_j(t_j) \geq 0 \quad for\ j=1,2,...,N \\ \sum_{j=1}^{N} \varepsilon_j(t_j) = \Pi \end{cases} \quad (4.7)$$

is a long-run equilibrium allocation.

Because of this, by Corollary 3.4 and Lemma 3.5, we immediately arrive at[11]:

---

[10] Strictly speaking, with each equilibrium revenue allocation one might associate several or many equilibrium production allocations. For example, we can not eliminate a possibility that there were another equilibrium production allocation $(y_1^a,...,y_N^a)$ whose every vector $y_j^a$ has two positive components: $y_{1j}^a$ and $y_{2j}^a$, so that $\varepsilon_j(t_j) = p_1 y_{1j}^a + p_2 y_{2j}^a$ for $j=1,...,N$, where $z(p) = \sum_{j=1}^{N} y_j^a$. Despite this, $(\varepsilon_1(t_1),...,\varepsilon_N(t_N))$ is still an equilibrium allocation since $(y_1^a,...,y_N^a)$ is an equilibrium allocation.

[11] When we here say that an equilibrium revenue allocation is Pareto optimal, we actually mean that the corresponding equilibrium production allocation is Pareto optimal. In this case, $(\varepsilon_1(t_1),...,\varepsilon_N(t_N))$ corresponds to $(y_1'(t_1),...,y_N'(t_N))$ at least, see (3.2) and (4.2).

***Corollary 4.3***: Any revenue allocation $(\varepsilon_1(t_1),...,\varepsilon_N(t_N))$ satisfying (4.7) is Pareto optimal.

Undoubtedly, (4.7) implies that there are infinitely many possible equilibrium outcomes. We later show that each equilibrium outcome $(\varepsilon_1(t_1),...,\varepsilon_N(t_N))$ can be depicted as a visual figure. *Temporarily*, we will drop the constraint $\sum_{j=1}^{N} \varepsilon_j(t_j) = \Pi$ in this subsection and the following two sections (concretely, subsection 4.2, sections 5 and subsections 6.1-6.2). However, we shall resume the constraint $\sum_{j=1}^{N} \varepsilon_j(t_j) = \Pi$ starting from subsection 6.3. Such a treatment will help readers to follow our idea more easily.

We now consider a simple LRCE with two firms as below:

***Example 4.1***: Assume that there exists a LRCE in which there are altogether two firms and two industries. Moreover, assume that if a firm enters the industry 1, then it will obtain $\varepsilon_1$ units of revenue; and that if a firm enters the industry 2, then it will obtain $\varepsilon_2$ units of revenue; and that $\varepsilon_1 < \varepsilon_2$.

Let us then explore how many equilibrium outcomes the Example 4.1 has. Because there are altogether two firms, we need to count all possible revenue allocations $(\varepsilon_1(t_1), \varepsilon_2(t_2))$ satisfying (4.7). However, because the constraint $\sum_{j=1}^{2} \varepsilon_j(t_j) = \Pi$ is *temporarily* dropped, we only need to count all possible revenue allocations $(\varepsilon_1(t_1), \varepsilon_2(t_2))$ satisfying $\varepsilon_j(t_j) \geq 0$ for $j = 1, 2$. As a consequence, there are altogether four equilibrium outcomes, which are respectively as follows:

$$A_1 = (\varepsilon_2, \varepsilon_2), \quad A_2 = (\varepsilon_1, \varepsilon_2), \quad A_3 = (\varepsilon_2, \varepsilon_1), \quad A_4 = (\varepsilon_1, \varepsilon_1).$$

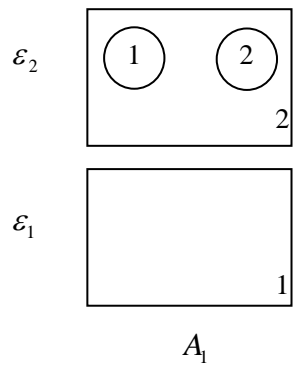

**Figure 1:** In the equilibrium outcome $A_1$, firms 1 and 2 both occupy the industry 2 and each obtains $\varepsilon_2$ units of revenue.

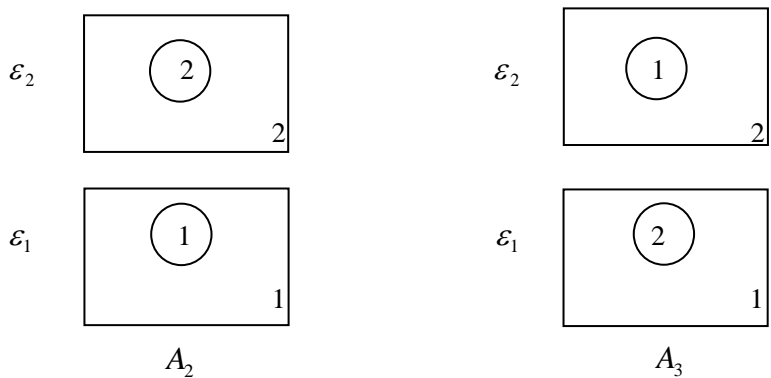

**Figure 2:** In the equilibrium outcome $A_2$, firm 1 occupies the industry 1 and obtains $\varepsilon_1$ units of revenue; firm 2 occupies the industry 2 and obtains $\varepsilon_2$ units of revenue. In the equilibrium outcome $A_3$, firm 1 occupies the industry 2 and obtains $\varepsilon_2$ units of revenue; firm 2 occupies the industry 1 and obtains $\varepsilon_1$ units of revenue.

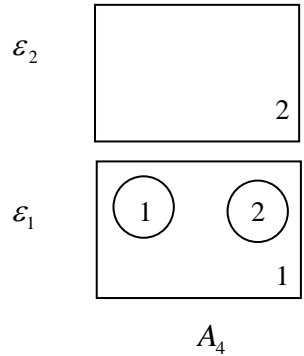

**Figure 3:** In the equilibrium outcome $A_4$, firms 1 and 2 both occupy the industry 1 and each obtains $\varepsilon_1$ units of revenue.

If we denote a firm by a ball and denote an industry by a box, then each equilibrium outcome $A_i$ $(i=1,2,3,4)$ can be depicted as a visual figure, see Figures 1-3. For example, the Figure 1 depicts the equilibrium outcome $A_1$ in which firms 1 and 2 both occupy the industry 2 and each thereby obtains $\varepsilon_2$ units of revenue, where ball 1 stands for firm 1 and box 1 stands for industry 1, and so forth.

Though the Example 4.1 merely describes a simple situation of (4.7) when $N=2$ and $\varepsilon_j(t_j)$ takes two possible values: $\varepsilon_1$ or $\varepsilon_2$, there are still four equilibrium outcomes. And because each equilibrium outcome is Pareto optimal, we are not able to make clear which equilibrium outcome is best for society so that all the social members would like to choose it. Of course, welfare economists think that one can find the best equilibrium outcome by taking advantage of an imaginary social welfare function. Unfortunately, *Arrow's Impossibility Theorem* has asserted that there is no such a social welfare function in the framework of ordinal utility (Jehle and Reny, 2001; Page 243). As a consequence, one would have to face an *uncertain* economic world which exhibits four possible social states[12]. However, this consequence seems to be inconsistent with the *scientific spirit of economics* which encourages the brave economists to explore the economic rule (or economic order) hidden behind the uncertain economic world. To proceed with this spirit, we attempt to extend the theoretical framework of Neoclassical economics by exhibiting Hayek's principle of spontaneous order. To arrive at this purpose, we next introduce the concept of economic order.

---

[12] From the viewpoint of empirical observation, there must be one and only one equilibrium outcome (or social state) which would occur (at a given time) even if we don't know which equilibrium outcome would occur.

## 5. Economic order

To simplify the analysis, we begin to introduce the concept of economic order by investigating the four equilibrium outcomes of the Example 4.1.

### 5.1. Definition

As pointed out in subsection 4.2, the LRCE described by Example 4.1 has four possible equilibrium outcomes: $A_1$, $A_2$, $A_3$ and $A_4$; each of which can be associated with a figure. If one observes the Figures 1-3 carefully, then one might find that these four outcomes can be divided into three different groups. To see this, we consider an ordered pair $\{a_1, a_2\}$, where $a_1$ represents that there are $a_1$ firms each of which obtains $\varepsilon_1$ units of revenue, and $a_2$ represents that there are $a_2$ firms each of which obtain $\varepsilon_2$ units of revenue. Adopting this notion one easily finds that the Figures 1-3 can be denoted by $\{a_1 = 0, a_2 = 2\}$, $\{a_1 = 1, a_2 = 1\}$ and $\{a_1 = 2, a_2 = 0\}$ respectively.

It is here worth mentioning that even if the Figure 2 depicts two equilibrium outcomes (thereby two figures): $A_2$ and $A_3$, we can still use a unique ordered pair $\{a_1 = 1, a_2 = 1\}$ to denote it. This is because $A_2$ and $A_3$ obey a unified rule or convention: One firm obtains $\varepsilon_1$ units of revenue and another obtains $\varepsilon_2$ units of revenue. From this reasoning, an ordered pair $\{a_1, a_2\}$ can be thought of as a 'set' whose elements are equilibrium outcomes.

For example, $A_2$ and $A_3$ obey the rule $\{a_1 = 1, a_2 = 1\}$, so we get:

$$\{a_1 = 1, a_2 = 1\} = \{A_2, A_3\}. \tag{5.1}$$

Likewise, we have:

$$\{a_1 = 0, a_2 = 2\} = \{A_1\}. \tag{5.2}$$

$$\{a_1 = 2, a_2 = 0\} = \{A_4\}. \tag{5.3}$$

If we extend the analysis about two firms above to $N$ firms, then we have the following definition about economic order.

***Definition 5.1*:** Let $W$ denote the set of all possible equilibrium outcomes satisfying (4.7). A sequence of non-negative numbers, $\{a_k\}_{k=1}^{n} = \{a_1, a_2, ..., a_n\}$, is called an economic order if and only if it denotes a subset of $W$ obeying the following four conventions:

(1). There are altogether $n$ possible revenue levels[13]: $\varepsilon_1 < \varepsilon_2 < ... < \varepsilon_n$;

(2). There are $a_k$ firms each of which obtains $\varepsilon_k$ units of revenue, and $k$ runs from 1 to $n$;

(3). These $a_k$ firms are distributed among $g_k$ industries[14];

---

[13] To guarantee that all possible equilibrium outcomes satisfying (4.7) can be appropriately (and without loss of any outcomes) divided into those different economic orders fulfilling Definition 5.1, we might require that $n \to \infty$ and $\varepsilon_{l+1} - \varepsilon_l \to 0$, where $l = 1, 2, ..., n-1$.

[14] It is worth mentioning that we can not prevent the possibility that $g_k > 1$. To see this, suppose that there were an equilibrium production allocation which contains several different equilibrium production vectors each of which generates a same revenue level. These different equilibrium production vectors (which must be linearly independent and otherwise can be thought of an industry) can be thought of as different industries. In this sense, however, (3.2) implies $g_k = 1$ for $k = 1, 2, ..., n$.

(4). $\sum_{k=1}^{n} a_k = N$.

It is easy to see that with every economic order $\{a_k\}_{k=1}^{n}$ one can associate a different revenue distribution as follow: There are $a_1$ firms each of which obtains $\varepsilon_1$ units of revenue, there are $a_2$ firms each of which obtain $\varepsilon_2$ units of revenue, and so on. From this meaning, an economic order actually denotes an ordered distribution rule of society's wealth. In general, any distribution rule is always due to some social institutions or conventions. In this sense, we are eager to make clear what distribution rule (or economic order) a free economy would obey. Hayek believed that if a competitive economy is enough fair and free, then a spontaneous economic order will arise. According to Hayek's thought, a striking feature of the spontaneous economic order is that it is *more likely* to emerge or more able to *survive* than other economic orders (Sugden, 1989). With this thought, we can present a concrete definition for spontaneous economic order as below:

***Definition 5.2***: For all possible economic orders $\{a_k\}_{k=1}^{n}$ satisfying Definition 5.1, if there exists an economic order $\{a_k^*\}_{k=1}^{n}$ which would occur with the highest probability, then $\{a_k^*\}_{k=1}^{n}$ is called a spontaneous economic order.

To understand the Definition 5.2, we can informally adopt the following statistical notion: One thinks of $W$ as a sample space in which each equilibrium outcome is regarded as a sample outcome (or an outcome of 'experiment'), and one thinks of an economic order as a random event which is identified with a collection of sample outcomes. Adopting such a notion, the spontaneous economic order is of course the most probable event (this is why it can arise spontaneously). In section 6, we shall formalize this notion and further show how to seek the spontaneous economic order from among all possible economic orders.

Before proceeding to do this, we are in particular interested in counting how many equilibrium outcomes a given economic order would contain. Let us next try to accomplish this task.

## 5.2. Monopolistic competition and perfect competition

For convenience, we here denote by[15] $\Omega(\{a_k\}_{k=1}^n)$ the number of elements in a given economic order $\{a_k\}_{k=1}^n$. Or equivalently we say that the economic order $\{a_k\}_{k=1}^n$ contains $\Omega(\{a_k\}_{k=1}^n)$ equilibrium outcomes. In microeconomics, we have made clear that there are two types of competitive structures, that is, perfect competition and monopolistic competition. Hence we need to find $\Omega(\{a_k\}_{k=1}^n)$ in terms of these two types of economic structures respectively.

***Definition 5.3*:** Monopolistic-competitive economy is an Arrow-Debreu economy in which firms are completely distinguishable[16] (or heterogeneous).

In fact, aimed at the LRCE described by Example 4.1, the Figures 1-3 have exhibited the situation of monopolistic competition where two balls (firms) are marked by serial numbers so that we can distinguish which is firm 1 and which is firm 2. For monopolistic-competitive LRCE, Tao (2010) has computed the number of elements in a given economic order $\{a_k\}_{k=1}^n$ in the form:

$$\Omega(\{a_k\}_{k=1}^n)_{mon} = \frac{N!}{\prod_{k=1}^n a_k!} \prod_{k=1}^n g_k^{a_k}. \tag{5.4}$$

For instance, the LRCE described by Example 4.1 requires that $N=2$, $n=2$ and $g_1 = g_2 = 1$. Thus, using the formula (5.4) we can compute the number of elements in each economic order as follows:

$$\Omega(\{a_1=0, a_2=2\})_{mon} = \frac{2!}{0! \times 2!} \times 1^0 \times 1^2 = 1, \tag{5.5}$$

$$\Omega(\{a_1=1, a_2=1\})_{mon} = \frac{2!}{1! \times 1!} \times 1^1 \times 1^1 = 2, \tag{5.6}$$

---

[15] Adopting this notation, $\Omega(\{a_k\}_{k=1}^n)$ should be a function of $a_k$, where $k=1,2,...,n$.
[16] Namely, that every firm corresponds to a different brand (Varian, 2003; Page 453)

$$\Omega(\{a_1 = 2, a_2 = 0\})_{mon} = \frac{2!}{2! \times 0!} \times 1^2 \times 1^0 = 1. \tag{5.7}$$

Clearly, the results (5.5)-(5.7) are consistent with the numbers of equilibrium outcomes listed by Figures 1-3 respectively.

***Definition 5.4***: Perfectly competitive economy is an Arrow-Debreu economy in which firms are indistinguishable[17] (or identical).

To see the difference between perfect competition and monopolistic competition, we continue to concentrate on the LRCE described by Example 4.1. Since firms are indistinguishable in perfectly competitive economy, we use the following three figures (see Figures 4-6) to list all possible equilibrium outcomes and the corresponding economic orders.

Typically, for the case of perfect competition, two balls (firms) are not marked by serial numbers (see Figures 4-6) so that we can not distinguish which is firm 1 or which is firm 2. Because of this, the economic order obeying perfect competition, $\{a_1 = 1, a_2 = 1\}$, contains only one equilibrium outcome, see Figure 5. This is clearly different from the case of monopolistic competition (comparing Figure 2 and Figure 5).

For perfectly competitive LRCE, Tao (2010) has computed the number of elements in a given economic order $\{a_k\}_{k=1}^n$ in the form:

$$\Omega(\{a_k\}_{k=1}^n)_{per} = \prod_{k=1}^n \frac{(a_k + g_k - 1)!}{a_k!(g_k - 1)!}. \tag{5.8}$$

---

[17] Namely, that firms produce homogeneous products (Varian, 2003; Page 380) so that the notion of brand does not exist. Perhaps, some people shall argue that homogeneous products, generally, only hold in one industry. However, in the long run, if a firm exits an industry, then it can enter an arbitrary industry in which there should not be differentiated products; otherwise there exists monopoly. As a result, homogeneous products, in the long run, hold in all industries; this case can be understood as products without brands (or equivalently, firms without brands).

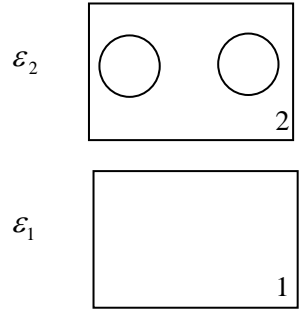

**Figure 4:** The economic order $\{a_1 = 0, a_2 = 2\}$ allows a single equilibrium outcome in which two indistinguishable firms occupy the industry 2 and each obtains $\varepsilon_2$ units of revenue.

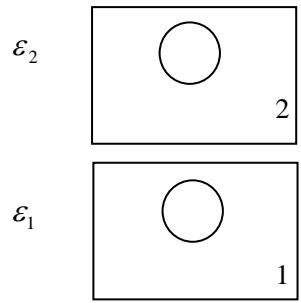

**Figure 5:** The economic order $\{a_1 = 1, a_2 = 1\}$ allows a single equilibrium outcome in which one firm occupies the industry 1 (hence obtains $\varepsilon_1$ units of revenue) and another firm occupies the industry 2 (hence obtains $\varepsilon_2$ units of revenue).

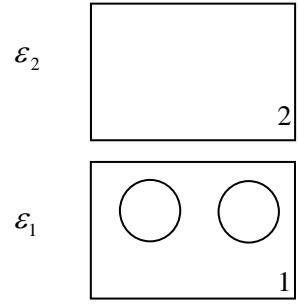

**Figure 6:** The economic order $\{a_1 = 2, a_2 = 0\}$ allows a single equilibrium outcome in which two indistinguishable firms occupy the industry 1 and each obtains $\varepsilon_1$ units of revenue.

Likewise, aimed at the LRCE described by Example 4.1, using the formula (5.8) we can compute the number of elements in each economic order as follows:

$$\Omega(\{a_1 = 0, a_2 = 2\})_{p\,e\,r} = \frac{(0+1-1)!}{0! \times (1-1)!} \times \frac{(2+1-1)!}{2! \times (1-1)!} = 1, \tag{5.9}$$

$$\Omega(\{a_1 = 1, a_2 = 1\})_{p\,e\,r} = \frac{(1+1-1)!}{1! \times (1-1)!} \times \frac{(1+1-1)!}{1! \times (1-1)!} = 1, \tag{5.10}$$

$$\Omega(\{a_1 = 2, a_2 = 0\})_{p\,e\,r} = \frac{(2+1-1)!}{2! \times (1-1)!} \times \frac{(0+1-1)!}{0! \times (1-1)!} = 1. \tag{5.11}$$

The results (5.9)-(5.11) are also consistent with the numbers of equilibrium outcomes listed by Figures 4-6 respectively.

In summary, we have:

$$\Omega(\{a_k\}_{k=1}^n) = \begin{cases} \prod_{k=1}^n \dfrac{(a_k + g_k - 1)!}{a_k!(g_k - 1)!} & (\text{perfect competition}) \\ \dfrac{N!}{\prod_{k=1}^n a_k!} \prod_{k=1}^n g_k^{a_k} & (\text{monopolistic } c\,o\,m\,p\,e\,tri) \end{cases} \tag{5.12}$$

# 6. Fairness, freedom and spontaneous economic order

In this section, we show how to seek the spontaneous economic order from among all possible economic orders by using the principles of fairness and freedom.

## 6.1. Social fairness

Undoubtedly, an ideal framework of thinking about social fairness or equity is the theory of social choice. To keep the following analysis simple we proceed to investigate the Example 4.1. As mentioned in subsection 4.2, the Example 4.1 has four possible equilibrium outcomes: $A_1$, $A_2$, $A_3$ and $A_4$; each of which is Pareto optimal. In this case, the task of the theory of social choice is to answer: Which of these four equilibrium outcomes is best for society. To accomplish this task, one may denote by $A = \{A_1, A_2, A_3, A_4\}$ the set of equilibrium outcomes. Then, if one can find some ranking of the equilibrium outcomes in $A$ that reflects 'society's' preferences, one would determine the best social choice. Unfortunately, *Arrow's Impossibility Theorem* has refuted the existence of such 'society's' preferences (Jehle and Reny, 2001; Page 243). Hence one is not able to compare any two alternatives in $A$ from a point of view which is individually consistent and social consistent; otherwise, the social choice will be unfair. To ensure fairness, a wise treatment is to abandon comparing any two equilibrium outcomes in $A$, and meanwhile to admit equality between all these equilibrium outcomes; that is,

$$A_1 \sim A_2 \sim A_3 \sim A_4, \qquad (6.1)$$

where, the symbol $\sim$ stands for the indifference relation.

Obviously, such a treatment is exhibiting Leibniz's *principle of the identity of indiscernibles* (Arrow, 1963; Page 109). Now that social members are indifferent between all equilibrium outcomes, we can not ensure which outcome will be selected as a collective decision. This means that collective choices should be completely random.

Because of the randomness of the collective choices, we are very interested in exploring the probability that a certain equilibrium outcome will be selected as a collective decision in a just society. To this end, let us concentrate on Rawls' *pure procedural justice* (Rawls, 1999; Page 74) which aims to design an economy (or economic institutions) so that the outcome is just whatever it happens to be, at least so long as it is within a certain range. With this idea, a just economy can be regarded as a fair procedure which will translate its fairness to the (equilibrium) outcomes, so that every social member would have no desire of opposing or preferring a certain outcome. That is, (6.1) holds. Technically, to insure that the economy is one of pure procedural justice, Rawls suggested to take into account the *principle of fair equality of opportunity* (Rawls, 1999; Page 76). Following this principle, a fair economy implies that each outcome should be selected with equal opportunities[18]; in other words, then each outcome will occur with an equal probability. Based on the analyses above, we can present the following axiom for social fairness.

*Axiom 6.1*: If a competitive economy produces $\omega$ equilibrium outcomes, and if this economy is absolutely fair, then each equilibrium outcome will occur with an equal probability $\frac{1}{\omega}$.

For instance, if the LRCE described by Example 4.1 is absolutely fair, then the Axiom 6.1 implies:

$$P[A_1] = P[A_2] = P[A_3] = P[A_4] = \frac{1}{4}, \qquad (6.2)$$

where, we denote by $P[X]$ the probability that an equilibrium outcome $X$ occurs.

Thanks to Axiom 6.1, we might apply the concept of classical probability (see page 21 in Larsen and Max (2001)) on the set of equilibrium outcomes of the LRCE. Concretely, we adopt the following three conventions:

(i). *The set of all possible equilibrium outcomes satisfying (4.7), $W$, is referred to as the sample space.*

(ii). *Each element (or equilibrium outcome) of the sample space $W$ is referred to as a sample outcome.*

(iii). *Each economic order which is identified with a collection of sample outcomes is referred to as a random event.*

---

[18] Following Rawls (1999; Page 134), fairness here has been modeled as a demand for insurance. For more investigations concentrating on the relationship between random choice and fairness, see Broome (1984).

Clearly, adopting the conventions (i)-(iii), we are able to compute the probability that any economic order occurs, provided that all possible equilibrium outcomes had been found. For example,

by (6.2) and (5.1) one has

$$P[\{a_1 = 1, a_2 = 1\}] = P[\{A_2, A_3\}] = P[A_2] + P[A_3] = \frac{1}{2},$$

and by (6.2) and (5.2) one has $P[\{a_1 = 0, a_2 = 2\}] = P[\{A_1\}] = P[A_1] = \frac{1}{4}$.

**6.2. Social freedom**

The concept of freedom is very complex, and every attempt to formalize it must neglect important aspects (Puppe, 1996). As most authors have done (Sen, 1993) (Pattanaik and Xu, 1998), this paper concentrates on the opportunity aspect of freedom. In this case, if social members are indifferent between alternatives, then the extent of freedom offered to the social members is entirely determined by the size of the set of alternatives (i.e., opportunity set), see Sen (1993).

In subsection 5.2, we have known that a given economic order $\{a_k\}_{k=1}^{n}$ contains $\Omega(\{a_k\}_{k=1}^{n})$ equilibrium outcomes. Therefore, if an economy obeys the economic order $\{a_k\}_{k=1}^{n}$, then the social members will face $\Omega(\{a_k\}_{k=1}^{n})$ possible choices. In the spirit of the opportunity-freedom (Sen, 1993), one can refer to an economic order as an opportunity set. Thanks to that social members are indifferent between equilibrium outcomes (see subsection 6.1), then the degree of freedom of an economic order might be denoted by its size (i.e., number of elements in it). Thus, we have the following axiom:

*Axiom 6.2*: If a competitive economy obeys an economic order $\{a_k\}_{k=1}^{n}$ which contains $\Omega(\{a_k\}_{k=1}^{n})$ equilibrium outcomes, then the degree of freedom of this economy is denoted by $\Omega(\{a_k\}_{k=1}^{n})$.

Since valuing freedom of choice might involve psychology (Verme, 2009), we shall not discuss the relationship between freedom and preference[19]. Also, we must here emphasize that[20] the degree of freedom defined by the Axiom 6.2 has no any ethical standard about 'good' or 'bad'. The larger degree of freedom merely implies the more possible choices. For example, by (5.1) the degree of freedom of $\{a_1 = 1, a_2 = 1\}$ equals 2, and by (5.2) the degree of freedom of $\{a_1 = 0, a_2 = 2\}$ equals 1. Then we do not mean that $\{a_1 = 1, a_2 = 1\}$ is better than $\{a_1 = 0, a_2 = 2\}$.

In subsection 5.1, we have defined the spontaneous economic order $\{a_k^*\}_{k=1}^n$ as an economic order with the highest probability, sees Definition 5.2. Following this definition, we shall now show that if a competitive economy not only is absolutely fair but also has the largest degree of freedom, then it would obey the spontaneous economic order $\{a_k^*\}_{k=1}^n$.

***Lemma 6.1***: If the competitive economy is absolutely fair, then the probability of the economic order $\{a_k\}_{k=1}^n$ occurring is given by:

$$P[\{a_k\}_{k=1}^n] = \frac{\Omega(\{a_k\}_{k=1}^n)}{\sum_{\{a_k'\}_{k=1}^n} \Omega(\{a_k'\}_{k=1}^n)}, \qquad (6.3)$$

where, $\sum_{\{a_k'\}_{k=1}^n} \Omega(\{a_k'\}_{k=1}^n)$ denotes the sum of the numbers of equilibrium outcomes over all possible economic orders.

---

[19] However, some authors believe that judgments about the degree of freedom offered to an agent by different opportunity sets must take into account the agent's preferences over alternatives, see Sen (1993), Kreps (1979) and Koopmans (1964).

[20] Sudgen (1998) also emphasized this point, and he further noted that the problem of measuring opportunity has many similarities with the familiar preference-aggregation problems of welfare economics and social choice theory.

***Proof.*** Because an economic order $\{a_k\}_{k=1}^n$ contains $\Omega\left(\{a_k\}_{k=1}^n\right)$ equilibrium outcomes, one easily counts that the competitive economy produces $\sum_{\{a_k\}}\Omega\left(\{a_k'\}_{k=1}^n\right)$ equilibrium outcomes altogether. According to the Axiom 6.1, if the competitive economy is absolutely fair, then each equilibrium outcome will occur with an equal probability $\dfrac{1}{\sum_{\{a_k'\}_{k=1}^n}\Omega\left(\{a_k'\}_{k=1}^n\right)}$.

Following the conventions (i)-(iii), the probability of the economic order $\{a_k\}_{k=1}^n$ occurring is denoted by (6.3). □

Using the Lemma 6.1, we can prove the following important result.

***Proposition 6.1:*** If a competitive economy is absolutely fair, and if it obeys the spontaneous economic order $\{a_k^*\}_{k=1}^n$, then it would have the largest degree of freedom; that is,

$$\Omega\left(\{a_k^*\}_{k=1}^n\right) = \max_{\{a_k\}_{k=1}^n}\Omega\left(\{a_k\}_{k=1}^n\right). \tag{6.4}$$

***Proof.*** According to the Definition 5.2, the spontaneous economic order $\{a_k^*\}_{k=1}^n$ is the most probable economic order, so we have:

$$P\left[\{a_k^*\}_{k=1}^n\right] = \max_{\{a_k\}_{k=1}^n} P\left[\{a_k\}_{k=1}^n\right]. \tag{6.5}$$

Substituting (6.3) into (6.5) yields (6.4). □

The proof above implies a corollary as below:

***Corollary 6.1:*** If a competitive economy not only is absolutely fair but also has the largest degree of freedom, then it would obey a spontaneous economic order.

The Proposition 6.1 is a central result of this paper since it not only tells us that a spontaneous economic order is completely determined by fairness, freedom and competition, but also implies a method of seeking the spontaneous economic order. It is easy to understand that seeking the spontaneous economic order $\{a_k^*\}_{k=1}^n$ is equivalent to solving an extremum problem $\max_{\{a_k\}_{k=1}^n} \Omega(\{a_k\}_{k=1}^n)$. This is what we will accomplish in the next subsection.

### 6.3. Spontaneous economic order

When we introduce the concept of economic order in section 5, we drop the constraint $\sum_{j=1}^{N} \varepsilon_j(t_j) = \Pi$ in (4.7). Undoubtedly, such a treatment is not strict. To see this, let us now return to the Example 4.1. It is easy to see that if one assumes $\varepsilon_1 + \varepsilon_2 = \Pi$, then one does have $\varepsilon_1 + \varepsilon_1 < \Pi$ and $\varepsilon_2 + \varepsilon_2 > \Pi$. As a result, the economic orders $\{a_1 = 2, a_2 = 0\}$ and $\{a_1 = 0, a_2 = 2\}$ do not satisfy (4.7) so long as $\varepsilon_1 + \varepsilon_2 = \Pi$. Consequently, to rule out the economic orders transgressing (4.7), we must resume the constraint $\sum_{j=1}^{N} \varepsilon_j(t_j) = \Pi$.

Without loss of generality, all the economic orders obeying (4.7) must satisfy the following two constraints:

$$\sum_{k=1}^{n} a_k = N, \qquad (6.6)$$

$$\sum_{k=1}^{n} a_k \varepsilon_k = \Pi. \qquad (6.7)$$

It is easy to see that (6.7) is just the constraint $\sum_{j=1}^{N} \varepsilon_j(t_j) = \Pi$. Thus, seeking the spontaneous economic order $\{a_k^*\}_{k=1}^n$ from among all the possible economic orders obeying (4.7) is equivalent to solving the extremum problem as below:

$$\begin{cases} \max_{\{a_k\}_{k=1}^n} \Omega\left(\{a_k\}_{k=1}^n\right) \\ s.t. \quad N = \sum_{k=1}^n a_k \\ \quad \Pi = \sum_{k=1}^n a_k \varepsilon_k \end{cases} \quad (6.8)$$

where, $\Omega\left(\{a_k\}_{k=1}^n\right)$ is denoted by (5.12).

To solve the extremum problem (6.8), we need to introduce a lemma.

***Lemma 6.2:*** Let $U[\Omega] = \ln \Omega\left(\{a_k\}_{k=1}^n\right)$. If $U[\Omega]$ reaches the maximum value at $\{a_k^*\}_{k=1}^n$, then $\Omega\left(\{a_k\}_{k=1}^n\right)$ reaches the maximum value at $\{a_k^*\}_{k=1}^n$ as well.

***Proof.*** If one observes that $U[\Omega]$ is a monotonically increasing function of $\Omega\left(\{a_k\}_{k=1}^n\right)$, then one easily completes the proof. □

Using the Lemma 6.2, the extremum problem (6.8) is equivalent to the following extremum problem:

$$\begin{cases} \max_{\{a_k\}_{k=1}^n} \ln \Omega\left(\{a_k\}_{k=1}^n\right) \\ s.t. \quad N = \sum_{k=1}^n a_k \\ \quad \Pi = \sum_{k=1}^n a_k \varepsilon_k \end{cases} \quad (6.9)$$

Substituting (5.12) into (6.9) we obtain the spontaneous economic order of the LRCE in the form:

$$a_k^*(I) = \frac{g_k}{e^{\alpha+\beta\varepsilon_k} - I} \quad \begin{cases} I = 1 & (\text{perfect competition}) \\ I = 0 & (\text{monopolistic competition}) \end{cases}, \quad (6.10)$$

$k = 1, 2, \ldots n,$

where, $\alpha$ and $\beta$ are Lagrange multipliers. Detailed calculations see Appendixes A and B.

The spontaneous economic order (6.10) is a central result of this paper; it is called the *Bose-Einstein distribution* whenever $I=1$, and is called the *Boltzmann distribution* whenever $I=0$ (Carter, 2001). Such an economic order determines the following rule of revenue distribution:

There are $a_k^*(I)$ firms each of which obtains $\varepsilon_k$ units of revenue, and $k$ runs from 1 to $n$.

Obviously, $a_k^*(I=0)$ will decrease exponentially as $\varepsilon_k$ (or $k$) grows. This result might strongly imply revenue inequality. However, revenue inequality does not contradict our definition for social fairness (see Axiom 6.1). In fact, Axiom 6.1 merely indicates that each firm has an equal chance of occupying any possible revenue level. In this case, to obtain a high revenue, luck and effort are likewise important, see Alesina and Angeletos (2005), Alesina, Cozzi and Mantovan (2012).

In addition, $\alpha$ and $\beta$ in (6.10) are two indeterminate multipliers, and both can not be determined by the spontaneous order theory itself.

## 7. Empirical investigation to spontaneous economic order

To make clear the economic meanings of $\alpha$ and $\beta$, we need to introduce the *Neoclassical macroeconomics* (Romer, 2000; Page 120) in which the aggregate revenue[21] $\Pi$ is completely determined by labor $L$, capital $K$ and technological progress $T$; that is,

$$\Pi = L^x K^y T^z. \qquad (7.1)$$

---

[21] Strictly speaking, we should here take the aggregate production function $z_m(p)$ rather than the aggregate revenue function $\Pi$, as introduced by Romer (2000). However, (4.4) implies that there is no essential difference between $z_m(p)$ and $\Pi$ (except a constant factor $p_m$).

As is well known, an important role of firm is to collect labor and capital. Naturally, firm can be thought of being composed of labor and capital (Williamson and Winter, 1993), e.g. a unit of firm corresponds to a unit of labor and capital. Hence the total number of firms, $N$, can be written as a function with respect to labor $L$ and capital $K$; that is,

$$N = N(L^x K^y). \tag{7.2}$$

Using (7.2), (7.1) can be rewritten in the form:

$$\Pi = \Pi(N, T). \tag{7.3}$$

Complete differential of (7.3) yields:

$$d\Pi(N, T) = \mu dN + \theta dT, \tag{7.4}$$

where, $\mu = \dfrac{\partial \Pi}{\partial N}$ and $\theta = \dfrac{\partial \Pi}{\partial T}$.

$\mu$ and $\theta$ denote the marginal labor-capital return and the marginal technology return of an economy respectively.

Using (6.6), (6.7) and (6.10), Tao (2010) arrived at:

$$d\Pi = -\frac{\alpha}{\beta} dN + \frac{1}{\beta} d\left( \ln W - \alpha \frac{\partial \ln W}{\partial \alpha} - \beta \frac{\partial \ln W}{\partial \beta} \right), \tag{7.5}$$

where $W = W(\alpha, \beta) = \prod_{k=1}^{n} \left(1 - Ie^{-\alpha - \beta \varepsilon_k}\right)^{-\frac{g_k}{I}}$.

Remarkably, the differential aggregate revenue (7.4) (from Neoclassical economics) and the differential aggregate revenue (7.5) (from spontaneous order theory) yield the same functional form. This means that Neoclassical economics and Austrian economics will be compatible with each other so long as (7.4) equals (7.5).

Following the thought of unifying Neoclassical and Austrian theories, by comparing (7.4) and (7.5), we get:

$$\alpha = -\frac{\mu}{\lambda \theta}, \tag{7.6}$$

$$\beta = \frac{1}{\lambda \theta}, \tag{7.7}$$

$$T = \lambda \left( \ln W - \alpha \frac{\partial \ln W}{\partial \alpha} - \beta \frac{\partial \ln W}{\partial \beta} \right), \tag{7.8}$$

where $\lambda$ is a positive constant.

Substituting (7.6) and (7.7) into (6.10) yields a definite form:

$$a_k^*(I) = \frac{g_k}{e^{\frac{\varepsilon_k - \mu}{\lambda\theta}} - I} \quad \begin{cases} I = 1 & (perfect\ competition) \\ I = 0 & (monopolistic\ competition) \end{cases}, \quad (7.9)$$

$k = 1, 2, \ldots n$.

(7.9) has earlier been obtained by Tao (2010). It describes the firms' revenue distribution in an economy. If we assign each firm to a different agent, then (7.9) may also describe the income distribution of a society. An attractive thought is to test (7.9) by collecting firms' revenue data or individuals' income data. Interestingly, there had been some empirical evidences supporting (7.9), see Yakovenko and Rosser (2009), Kürten and Kusmartsev (2011), Kusmartsev (2011), Clementi et al (2012). Let us next introduce how these empirical investigations support (7.9).

It is easy to see that $\{a_k^*(I = 0)\}_{k=1}^n$ is an exponential distribution. Such a distribution is associated with the monopolistic-competitive economy. As pointed out in microeconomics, monopolistic competition is a common competitive mode, and most real economies usually obey this mode. Yakovenko and Rosser (2009) have confirmed that the income distribution in USA during from 1983 to 2000 well obeys the exponential distribution. Later, Clementi et al (2012) also confirmed this point. More concretely, Yakovenko and Rosser (2009) show that about 3% of the population obey Pareto distribution (i.e., power-law distribution), and 97% obey Boltzmann distribution (i.e., exponential distribution). This fact that income distribution consists of two distinct parts reveals the two-class structure of the American society[22].

However, $\{a_k^*(I = 1)\}_{k=1}^n$ is a unstable distribution. To see this, one only needs to notice that there may be a singularity $\varepsilon_k = \mu$ so that the denominator of (7.9) corresponding to $I = 1$ equals zero. Because of this, there might be many and many firms (or agents) occupying a very low revenue level (or income level) $\mu$, details see the section IV in Tao (2010).

---

[22] According to our theory, Boltzmann distribution is a consequence of free competition; hence, we infer that about 97% of the population in the American society obey the rule of free competition, but the remaining fraction might consist of monopolist (or privileged class).

Such a unstable distribution is associated with the case of extreme competition, i.e., perfect competition. Recently, Kürten and Kusmartsev (2011) have confirmed that the income distribution in USA between the years 1996-2008 well obeys this distribution; also, that the financial crisis in 2008 is due to the instability of this distribution.

## 8. Technological progress and freedom

In Neoclassical economics, the technological progress $T$ (sees (7.1)) is mysterious, and nobody makes clear what is the origin of it. Of course, there had been some excellent economic models, e.g. Romer (1990), in which the technological progress is interpreted as an endogenous variable, whereas it is artificially taken into these models. Remarkably, soon we shall see that if one treats Neoclassical economics and Austrian economics in a unified manner, one will decipher the profound origin of technological progress.

Using (5.12) and (6.10), Tao (2010) arrived at:

$$\ln \Omega(\{a_k^*\}) = \ln W - \alpha \frac{\partial \ln W}{\partial \alpha} - \beta \frac{\partial \ln W}{\partial \beta}. \qquad (8.1)$$

Substituting (8.1) into (7.8) yields a refined form[23]:
$$T = \lambda \ln \Omega. \qquad (8.2)$$

From (8.2), we surprisingly find that the technological progress $T$ is exactly proportional to $\ln \Omega$. Also, because $\Omega$ (or equivalently $\ln \Omega$) stands for the degree of freedom of an economy, whence we conclude: The more freedom, and the more rapid technological progress.

We try to convey an intuition for the result above. To make things simple we look at the Figures 1 and 2. The Figure 2 depicts the economic order $\{a_1 = 1, a_2 = 1\}$ whose degree of freedom is denoted by 2, then firm 1 not only may enter industry 1 but also may enter industry 2. By contrast, the degree of freedom of $\{a_1 = 0, a_2 = 2\}$ is denoted by 1, then firm 1 is confined within the industry 2 (see Figure 1). Logically, if a firm has the chance of entering two industries, then the probability of causing innovation should increase relative to only being confined within one industry. That is to say, the more freedom (namely, the larger $\Omega$), the larger probability of causing innovation.

---

[23] (8.2) implies that technological progress $T$ looks like the entropy in physics (Tao, 2010). Interestingly, the latter is often related to "information" or "knowledge".

As is well known, Schumpeter (1934) ever emphasized that innovation is a main driving force of promoting economic development. In the other direction, Hayek (1948) believed truly that freedom will induce a spontaneous economic order so that the economic development is most efficient. Interestingly, (8.2) undoubtedly indicates that the innovation emphasized by Schumpeter is essentially equivalent to the freedom highlighted by Hayek. In other words, our theory has unified the ideas of Schumpeter and Hayek which seem independent each other. From this meaning, when an economy favors a state with more freedom, it essentially favors a state (or evolutionary direction) with higher technology level as well. It is worth mentioning that the "freedom" is a natural endogenous variable in our theory[24]. To see this, substituting (8.2) into (7.3) we obtain:

$$\Pi = \hat{\Pi}(N, \Omega). \qquad (8.3)$$

From (8.3), we see that the freedom ($\Omega$), as an equivalent replacement of technological progress ($T$), will become a possible driving force of promoting economic growth. Actually, some empirical studies have found a non-linear relationship between economic freedom and growth, see Barro (1996).

So far, we have presented a complete theoretical framework for spontaneous economic order. At the moment, we need to review the logical setups of this theoretical framework. First, we prove that a LRCE does produce infinite many equilibrium outcomes. Second, we show that all these equilibrium outcomes can be appropriately (non-repeated) assigned into some economic orders. Third, according to the principles of fairness and freedom, one can determine an economic order which will occur with the highest probability, and we dub such an economic order the spontaneous economic order. Finally, we verify that Austrian economics and Neoclassical economics will become compatible with each other within the framework of spontaneous economic order, provided that the technological progress $T$ is proportional to the freedom variable $\ln \Omega$. Because of these above, we believe that the spontaneous order theory presents a possible link between Austrian economics and Neoclassical economics.

## 9. Conclusion

This paper presents a theoretical framework for spontaneous economic order in which the union of Austrian economics and Neoclassical economics lies at the heart of our attempt. Our study shows that if a competitive economy is enough fair and free, then an optimal economic order shall emerge spontaneously. It is worthwhile to note that such an economic order is not the result of any process of collective choice (unlike expected by many welfare economists), but is an unplanned and spontaneous consequence (as expected by Hayek).

---

[24] Because of this, technological progress $T$ is also an endogenous variable.

Generally speaking, we can not guarantee that an Arrow-Debreu economy has only one equilibrium outcome. If an Arrow-Debreu economy (e.g., long-run competitive situation) produces multiple equilibrium outcomes, then according to the first fundamental theorem of Welfare economics each outcome will be Pareto optimal. Consequently, social members will have to face a problem of social choice: They must choose an equilibrium outcome which is best for society. For this problem, the proposal of welfare economists is to search the best outcome through an imaginary social welfare function. Nevertheless, *Arrow's Impossibility Theorem* warns us: There will be no such a social welfare function if we insist on the standpoint of ordinal utility. This means, a plan of seeking the best equilibrium outcome through an social welfare function is doomed to failure.

Our proposal here is to abandon searching the best equilibrium outcome and meanwhile to admit the equality between all possible equilibrium outcomes. In this case, we try to divide all these equilibrium outcomes into some groups each of which exhibits a different convention. These conventions are called the economic orders by us. Based on these preparations above, we show that if one adds some normative criteria about fairness and freedom into the Arrow-Debreu economy, one will find that there does exist an economic order with the highest probability which we call the spontaneous economic order. Undoubtedly, an economic order with the highest probability would *most likely* occur, this is why we call it the *spontaneous* order. In this sense, we can say that the economic world does change the way it does because it seeks an economic order of higher probability. Our attempt has very strong theoretical and practical significance: The goal of human society should be to insist on the criteria about fairness and freedom (similar to Axioms 6.1-6.2). Following these criteria, the competitive society will automatically obey a spontaneous economic order. Concretely, we conclude that the spontaneous order of a monopolistic-competitive economy will obey a stable rule: Boltzmann distribution; and that the spontaneous order of a perfectly competitive economy will obey an unstable rule: Bose-Einstein distribution. And the instability of the latter may cause economic crises. Our these conclusions have been supported by some recent empirical investigations.

Obviously, our spontaneous order theory is in principle on the basis of the theoretical framework of Arrow-Debreu economy, so it may present a bridge linking Austrian economics and Neoclassical economics. An interesting conjecture is: Might Austrian economics and Neoclassical economics constitute a unified framework? Our spontaneous order theory confirms this conjecture, provided that the technological progress in Neoclassical economics and the "freedom" in Austrian economics become equivalent with each other. As an application of unifying these two types of economics, we shall comprehend a truth: "Freedom promotes technological progress."

## Appendixes

**A. Spontaneous economic order of perfectly competitive economy**

Allow for that the number of firms $N \to \infty$ in a long-run competitive economy, we assume that every $a_k$ is a sufficiently large number.

If one considers the perfect competition, then using (5.12) the function $U[\Omega]$ can be written in the form:

$$U[\Omega_{pe}] = \sum_{k=1}^{n} \ln(a_k + g_k - 1)! - \sum_{k=1}^{n} \ln a_k! - \sum_{k=1}^{n} \ln(g_k - 1)!. \quad (A.1)$$

Thanks to that the values of $a_k$ and $g_k$ are large enough, using the Stirling's formula (Carter, 2001; Page 218)

$$\ln m! = m(\ln m - 1), \quad (m >> 1) \quad (A.2)$$

(A.1) can be rewritten in the form:

$$U[\Omega_{pe}] = \sum_{k=1}^{n} [(a_k + g_k - 1)\ln(a_k + g_k - 1) - a_k \ln a_k - (g_k - 1)\ln(g_k - 1)]. \quad (A.3)$$

The method of Lagrange multiplier for the optimal problem (6.9) gives

$$\frac{\partial \{U[\Omega]\}}{\partial a_k} - \alpha \frac{\partial N}{\partial a_k} - \beta \frac{\partial \Pi}{\partial a_k} = 0, \quad k = 1, 2, ..., n \quad (A.4)$$

where, $\alpha$ and $\beta$ are Lagrange multipliers.

Substituting (6.6), (6.7) and (A.3) into (A.4) yields

$$\ln\left(\frac{a_k + g_k}{a_k} - \alpha - \beta\varepsilon_k\right) = 0, \qquad (A.5)$$

$k = 1,2,\ldots,n$.

which is the spontaneous economic order of perfectly competitive economy:

$$a_k = \frac{g_k}{e^{\alpha+\beta\varepsilon_k} - 1}, \qquad (A.6)$$

$k = 1,2,\ldots,n$.

## B. Spontaneous economic order of monopolistic-competitive economy

If one considers the monopolistic competition, then using (5.12) the function $U[\Omega]$ can be written in the form:

$$U[\Omega_{mon}] = \ln N! + \sum_{k=1}^{n} a_k \ln g_k - \sum_{k=1}^{n} \ln a_k!. \qquad (B.1)$$

Using the Stirling's formula (A.2), (B.1) can be rewritten in the form:

$$U[\Omega_{mon}] = \ln N! + \sum_{k=1}^{n} a_k \ln g_k - \sum_{k=1}^{n} a_k \ln a_k + \sum_{k=1}^{n} a_k. \qquad (B.2)$$

Substituting (6.6), (6.7) and (B.2) into (A.4) gives the spontaneous economic order of monopolistic-competitive economy:

$$a_k = \frac{g_k}{e^{\alpha+\beta\varepsilon_k}}, \qquad (B.3)$$

$k = 1,2,\ldots n$.